\documentstyle[hx37_latex]{article}

\def\mco{\multicolumn}

\def\be{\begin{equation}}
\def\ee{\end{equation}}
\def\bea{\begin{eqnarray}}
\def\eea{\end{eqnarray}}

\begin{document}

\title{OLD GALAXIES AT HIGH REDSHIFT}

\author{J.S. DUNLOP}

\address{Institute for Astronomy, Department of Physics \& Astronomy,
University of Edinburgh, Royal Observatory, Edinburgh EH9 3HJ, UK}

\maketitle\abstracts{
The most passive galaxies at high redshift are unlikely to be identified
by either narrow-band emission-line searches, or by Lyman limit searches
(both techniques which have been highlighted at this meeting) simply
because such selection methods rely on the presence of a strong
ultraviolet component. 
Selection on the basis of extreme radio power has
also proved to yield optically active objects with the majority of
high-redshift objects studied to date displaying complex elongated
optical/UV morphologies, relatively blue optical-ultraviolet continuum
colours, and strong emission lines. These features, coupled with the
failure to detect any spectral signatures of old stars at $z > 1$, has
led to the suggestion that these galaxies are being observed close to or
even during a general epoch of formation. However, we have recently
demonstrated that radio selection at significantly fainter (mJy) flux
densities can be used to identify apparently passively evolving elliptical
galaxies at high redshift. Deep Keck spectra have now been obtained for two
such objects yielding absorption line redshifts $z \simeq 1.5$; 53W091 at $z = 1.552$ (Dunlop {\it
et al.} 1996)$^1$ and most recently 53W069 at $z  = 1.432$. The ultraviolet
SEDs of these galaxies indicate minimum ages $> 3$ Gyr while, as stressed in
this article, the strength of the reddenning-independent ultraviolet
spectral breaks actually indicate a greater 
minimum age of 5 Gyr for both objects assuming
solar metallicity. Since the spectra comprise the 
integrated light of each galaxy to radii greater than $r_e$, 
I argue that it is difficult
to justify the adoption of significantly super-solar metallicity 
in interpreting these data. It thus seems hard to escape
the conclusion that $\Omega_0 < 1$ and that, irrespective of the adopted
cosmology, at least some 
massive ellipitical galaxies were formed at high redshift ($z
> 5$).}

\section{Background: locating passively evolving galaxies at high z}

The recent discovery of a substantial population of star-forming galaxies
at ($3.0 < z < 3.5$) has revolutionised the study of radio quiet galaxies
at high redshift, as evidenced by a number of contributions at this
meeting$^{2}$. 
However a selection method which depends 
on a Lyman continuum break superimposed on an otherwise blue far-UV 
continuum can shed little
light on the evolutionary state of the most passively evolving
systems which exist at a given epoch. This is unfortunate since, given
the ease with which a relatively small starburst can mask the true
properties of an underlying galaxy, it is the reddest/most-passive
systems at any redshift which are of greatest interest for constraining
the first epoch of galaxy formation and indeed the age of the Universe.

Radio-based selection has long provided an alternative and 
effective method of locating high
redshift galaxies which, at least in principle, should not be so directly 
biassed towards star-forming sources. Indeed, if anything it should be
biassed towards the precursors of old elliptical galaxies since at 
low-redshifts it is well-established that the hosts of powerful radio sources 
are elliptical galaxies with well-evolved stellar
populations.
Despite this, identification of high-redshift objects 
on the basis of extreme radio power has
also yielded optically active objects with the majority of
high-redshift radio galaxies studied to date displaying complex elongated
optical/UV morphologies, relatively blue optical-ultraviolet continuum
colours, and strong emission lines$^{3}$. However, the fact that the
optical-ultraviolet properties of high-redshift radio galaxies are known
to correlate with radio power$^4$ suggests that any radio-based search for `normal'
elliptical galaxies at high redshift should be confined to milli-Jansky
flux levels. Accordingly, over the past few years we have 
investigated the properties of weak radio galaxies with $S_{1.4GHz} > 1$
mJy from the Leiden Berkeley Deep Survey, and have isolated a
sample of 10 extremely red objects that have $R-K > 5$ and 
$z_{est} > 1$ for intensive spectroscopic study. 

\section{Keck Spectroscopy}

While a red $R-K$ colour can be taken as indicative of an old stellar
population, deep optical spectroscopy is vital for the reliable dating of
these objects for four reasons. First, a spectroscopic redshift is
required. Second, it is necessary to show that the red
colour of the object arises from a lack of young stars rather than, for
example, from a dust-reddened active nucleus. Third, the 
shape of the rest-frame 
ultraviolet spectrum of a galaxy is extremely sensitive to the age of the
stellar population$^{5}$. Fourth, for high-redshift galaxies 
it should be possible to use evolutionary
synthesis models to derive relatively robust age estimates from 
ultra-violet SEDs because, for the potential age range of interest 
({\it i.e.} age $< 8$ Gyr for $z > 1$) the ultraviolet SED is completely
dominated by stars near the main-sequence turnoff point on the HR
diagram$^{6}$
({\it i.e.} disagreements over, for example, the strength and colour of
the AGB or HB are unimportant).

\begin{figure}
\setlength{\unitlength}{1mm}
\begin{picture}(95,70)
\includegraphics{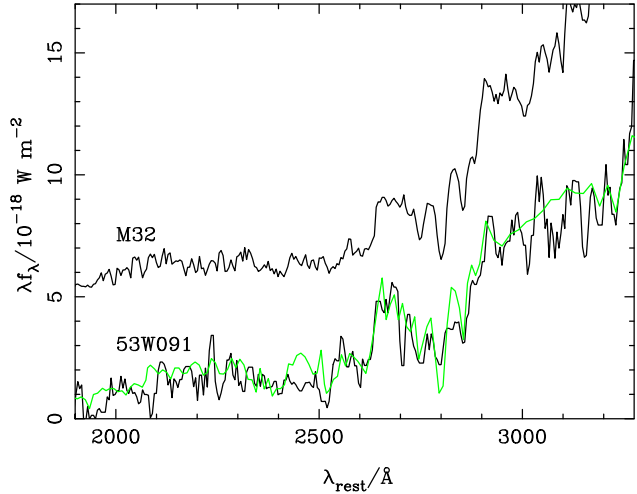}
\end{picture}
\caption{The rest-frame spectrum of 53W091 compared with an
instantaneous starburst at an age of 3.5 Gyr (dotted line) and a transposed
IUE spectrum of the elliptical M32.}
\label{fig1}
\end{figure}

We have now obtained deep optical spectra of two 
red mJy radio galaxies using LRIS on the Keck telescope. 
Our redshift determination and spectral dating of the first of 
these (53W091; $z = 1.552$) have been published$^1$
and will be described in more detail elsewhere$^7$. In brief, the
ultraviolet SED of this source is, as illustrated in Figure 1, 
very similar to those of low-redshift ellipticals such as M32, and 
essentially identical to that of an F6V star. Both these comparisons
suggest an age of $\simeq 3.5$ Gyr, a result confirmed by spectral
synthesis modelling.
We have recently (June 1996) obtained a deep LRIS spectrum of a second red mJy radio
galaxy, 53W069. This object also lies at $z \simeq 1.5$ ($z = 1.432$) and
has an ultraviolet SED which is in fact slightly redder, indicating an
age of $\simeq 4.5$ Gyr$^{8}$.

\section{Galaxy ages from ultraviolet spectral breaks }

While the large number of stellar absorption features detected in the
spectra of both these objects proves that their UV light is dominated by
stars, dating on the basis of the overall shape of their UV SED is
susceptible to distortion either by dust reddening or by low-level 
direct/indirect AGN contamination. However the strengths of the spectral
breaks at 2640\AA\ and 2900\AA\, being relatively immune from such
complications should yield more robust age
estimates. Certainly the strength of these breaks in the IUE spectra of stars 
is well studied$^{9}$, and despite misgivings over the understanding of the
relevant opacities, evolutionary synthesis models based on both observed
and theoretical stellar spectra do indeed seem to produce reasonably
consistent results from both breaks. As indicated in
Table 1, this analysis indicates that in both galaxies $\simeq 5$ Gyr has
elapsed since the last era of significant star-formation activity.

The most attractive way to reconcile an Einstein-de Sitter Universe with 
such large ages at $z \simeq 1.5$ is to assume that the strong breaks in 
both 53W069 and 53W091 are due to high metallicity rather than
age. However, the metallicity dependence of these breaks does not
appear to be strong$^{1,9}$, and in any case the {\it mean} metallicities of
comparably massive giant ellipticals at low redshift are at most only mildly
super-solar when averaged out to $r \simeq r_e$ ($\simeq 0.5$ arcsec at
$z \simeq 1.5$ and thus within the LRIS slit)$^{10}$.
 Taken at face value an age 
$> 5$ Gyr at $z \simeq 1.5$ implies $\Omega_0 < 0.2$ for $H_0 > 55 {\rm km
s^{-1} Mpc^{-1}}$ unless $\Lambda > 0$.

\begin{table}[t]
\caption{Minimum ages deduced from the break strengths in the rest-frame
spectra of 53W091 and 53W069 using 3 alternative models of galaxy spectral 
evolution. The evolution of the 2640\AA\ break in Worthey's model 
is anomolously rapid ({\it e.g.} it yields a 
MS turnoff age of $< 3$ Gyr for the sun). The other
ages in the table are consistent with 5 Gyr.}
\vspace{0.4cm}
\begin{center}
\begin{tabular}{|c|c|c|c|}
\hline
\mco{2}{|c|}{Ages (Gyr)} &Model &Average Age \\
\hline
2640\AA & 2900\AA & & \\
\cline{1-2}
(1.8)     &   4.6   & Worthey  (1994)$^{11}$ &  4.6\\
4.0  &   6.5   & Jimenez {\it et al.} (1996)$^{12}$  &  5.3\\
6.2  &   4.6   & Bruzual \& Charlot (1993)$^{13}$    &  5.4\\ \hline
\end{tabular}
\end{center}
\end{table}

\section*{Acknowledgments}
I gratefully acknowledge my collaborators in this work: 
John Peacock, Hy Spinrad, Arjun
Dey, Rogier Windhorst, Raul Jimenez and Daniel Stern.

\end{document}